\documentclass[namedreferences]{solarphysics}

\usepackage[hyperref,optionalrh]{spr-sola-addons} 
\usepackage{graphicx}        
\usepackage{color}           
\usepackage{breakurl}        

\usepackage{natbib}
\usepackage{multirow}

\begin{document}

\begin{article}
\begin{opening}

\title{Robustness of Solar-Cycle Empirical Rules Across Different Series Including an Updated ADF Sunspot Group Series }

\author[addressref={aff1,aff2},corref,email={ilya.usoskin@oulu.fi}]{\inits{I.}\fnm{Ilya}~\lnm{Usoskin}\orcid{0000-0001-8227-9081}}
\author[addressref=aff3]{\inits{G.A.}\fnm{Gennady}~\lnm{Kovaltsov}\orcid{0000-0002-4202-1032}}
\author[addressref={aff1,aff4}]{\inits{W.}\fnm{Wilma}~\lnm{Kiviaho}\orcid{0000-0001-6690-6540}}
\address[id=aff1]{University of Oulu, Finland}
\address[id=aff2]{St.Petersburg State University, Russia}
\address[id=aff3]{A.F. Ioffe Physical-Technical Institute, St. Petersburg, Russia}
\address[id=aff4]{University of Turku, Finland}

\runningauthor{I. Usoskin et al.}
\runningtitle{Cycle Empirical Rules}

\begin{abstract}
Empirical rules of solar cycle evolution form important observational constraints for the solar dynamo theory.
This includes the \textit{Waldmeier rule} relating the magnitude of a solar cycle to the length of its ascending phase, and the
 \textit{Gnevyshev--Ohl rule} clustering cycles to pairs of an even-numbered cycle followed by a stronger odd-numbered cycle.
These rules were established as based on the ``classical'' Wolf sunspot number series, which has been essentially revisited recently,
 with several revised sets released by the research community.
Here we test the robustness of these empirical rules for different sunspot (group) series for the period 1749\,--\,1996, using
 four classical and revised international sunspot numbers and group sunspot-number series.
We also provide an update of the sunspot group series based on the active-day fraction (ADF) method, using the new
 database of solar observations.
We show that the Waldmeier rule is robust and independent of the exact sunspot (group) series:
 its classical and $n+1$ (relating the length of $n$-th cycle to the magnitude of ($n+1$)-th cycle) formulations are significant
 or highly significant for all series, while its
 simplified formulation (relating the magnitude of a cycle to its full length) is insignificant for all series.
The Gnevyshev--Ohl rule was found robust for all analyzed series for Cycles 8\,--\,21, but unstable across the Dalton minimum and before it.
\end{abstract}
\end{opening}

\section{Introduction}
\label{S:Intro}

The series of sunspot number is the longest and probably the most analyzed astrophysical dataset based on
 consistent direct telescopic observations since 1610 \citep[e.g.][]{clette14,usoskin_LR_17}.
It provides a quantitative index of solar magnetic variability, which is dominated by the $\approx$11-year
 \textit{Schwabe} cycle.
However, this cyclic variability also exhibits variations of individual cycles both in strength/magnitude
 and the length \citep[see a detailed review by][]{hathawayLR}.
This includes also the Maunder minimum (1645\,--\,1715) when sunspots activity was dramatically suppressed \citep{eddy76,usoskin_MM_15}
 and the Modern grand maximum in the second half of the 20th century \citep{solanki_Nat_04,lockwood_JSWSC_18}.

Because of the length of the sunspot series, solar cyclic variability has been extensively studied including several
 empirical rules relating the temporal evolution of the solar cycles with their strengths.
Here we consider the following empirical rules.
One is the \textit{Waldmeier rule}  \citep{waldmeier35,waldmeier39} which relates the duration of the ascending phase of a solar cycle
 to its amplitude.
The other is the \textit{Gnevyshev--Ohl} rule \citep{gnevyshev48} (sometimes called the Even--Odd effect), which implies a
 strong connection between the integral intensities of an even-numbered and the subsequent odd-numbered cycles.
These rules are empirical, they are based on an analysis of the data and they serve as important observational constraints for
 solar dynamo models \citep[e.g.][]{charbonneauLR}.

For more than a century, the sunspot series, initially compiled by Rudolf Wolf and continued by his successors in Z\"urich,
 called the \textit{Wolf sunspot series} (WSN), remained the ``golden truth'' forming a basis for numerous studies.
The WSN series is composed via a linearly scaled combination of the counts of sunspot groups and individual spots
 reported by different observers with different skills and instrumentation \citep[see details in][]{hathawayLR}.
This WSN series was continued as the \textit{International sunspot series} (ISN) by the Royal Observatory of Belgium since 1981
 \citep{clette07}.
However, the approach of R. Wolf in compiling the WSN was questioned by \citet{hoyt98}, who proposed that the use of only
 the number of sunspot groups is more robust and revised the entire set of the individual observers' data, proposing an
 alternative \textit{Group sunspot number} (GSN).
It was shown by \citet{hathaway02} that the use of the GSN series slightly alters the empirical relations obtained for the WSN/ISN series.

As demonstrated several years ago, the ``classical'' WSN and GSN series contain a number of inconsistencies or errors that compromise
 the reliability of the data before the onset of the 20th century \citep[e.g.][]{clette14,usoskin_LR_17,munos19}.
Moreover, new data of historical observations have been found and analyzed
 \citep[e.g.][]{lockwood_1_14,vaquero_rev07,arlt13,arltLR,hayakawa_derfflinger_20,carrasco20}.
As a result, several updates and revisions of the sunspot \citep{clette16,chatzistergos17} and group
 \citep{cliver16,svalgaard16,usoskin_ADF_16,willamo17} number series have been recently provided.

The empirical rules were studied by different authors
 \citep[e.g.][]{usoskin_lost_AA_01,usoskin_lost_GRL_02,hathaway02,aparicio12,tlatov13,hathawayLR,carrasco_cycle_16,takalo18}
 using different sunspot series, WSN and GSN, and they were not tested systematically.

Thus, there is a need to revisit the empirical rules based on the sunspot series and check their robustness against the
 exact dataset used.
This forms the main focus of this work.

\section{Sunspot Series}
\label{S:data}
\begin{figure}[t]
\centering
\includegraphics[width=0.8\columnwidth]{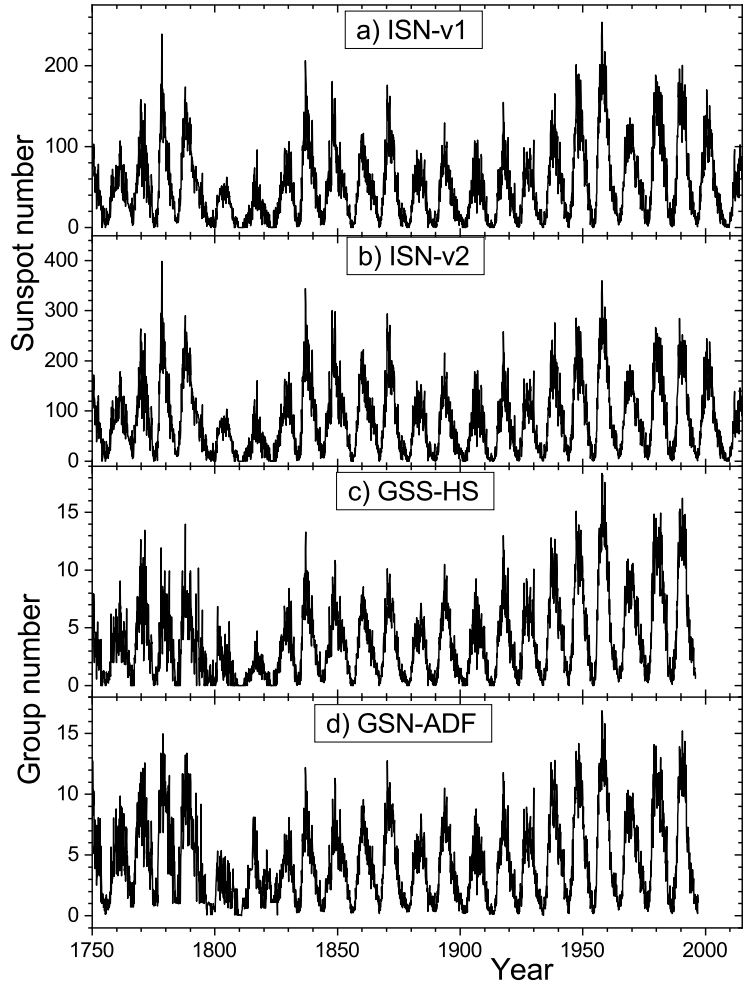}
\caption{Time series of the sunspot (group) numbers used as described in Section~\ref{S:data}.}
\label{Fig:data}
\end{figure}

For the analysis we selected sunspot series that fulfil the following criteria:
\begin{itemize}
\item
The series should extend back until 1749, viz. being of the same length as the original WSN.
Although some series extend back to 1610, we do not consider the period before 1749 because of the very uncertain
 data between 1720\,--\,1740 and the grand-minimum state of solar activity during the Maunder minimum (1645\,--\,1715).
\item
The series should have at least monthly resolution.
Annual resolution is insufficient to analyze the Waldmeier rule.
\end{itemize}
Four series satisfy these criteria.
These series are shown in Figure~\ref{Fig:data}.
\begin{enumerate}
\item
The ``classical'' WSN, which was the basis for most of the previous analyzes of the empirical rules, was taken
 as a monthly series for January 1749 through May 2015 according to the ISN version 1.0 dataset.
It was taken from SILSO (www.sidc.be/silso/datafiles) and is referred further as \textit{ISN-v1}.

\item
The revised ISN includes recent updates and corrections and is referred to as ISN version 2.0 \citep{clette16}.
Its was taken, as a monthly series for January 1749 through August 2020, from SILSO and is referred further as \textit{ISN-v2}.

\item
The `classical' GSN series by \citet{hoyt98} is further referred as \textit{GSN-HS}.
We used the monthly series covering the time from January 1749 through December 1995, as downloaded from SILSO.

\item
A new method of composing the GSN series, called the Active Day Fraction (ADF) method, was proposed
 recently \citep{usoskin_ADF_16,willamo17}.
Here we have updated it by exploiting the renewed database of individual observer's data
 (haso.unex.es/haso/index.php/on-line-archive/data/) \citep{vaquero16}.
The details of the update are presented in Appendix~\ref{App:A}.
It is hereafter referenced as the \textit{GSN-ADF} series.
This new series supersedes the previous ADF series \citep{usoskin_ADF_16,willamo17}.
\end{enumerate}
All series were reduced to the overlapping period including Cycles 1\,--\,22 (1755--1996) for the sake of consistency.
Some series contain data gaps in the earlier part.
The gaps shorter than seven months were interpolated by 13-month running mean values,
 while two longer gaps in the GSN-ADF series (19 months in June 1810\,--\,December 1811, and 17 months in November 
 1814\,--\,March 1816) were interpolated using a 41-month parabolic fit \citep[similar to][]{mursula01}.

Another type of GSN reconstructions, based on the original \citep{svalgaard16} or improved \citep{chatzistergos17} ``backbone'' method,
 provide only annual resolution and are not analyzed here.

\section{Cycle Parameters}
\label{S:cycle}
Here we are interested in such parameters of the solar cycle as its phasing (ascending, descending phase and total length),
 magnitude, and intensity (cumulative sum of sunspot number over the cycle).

While some community-accepted dates of solar-cycle minima and maxima are listed for WSN
 (ftp.ngdc.noaa.gov/space-weather/solar-data/solar-indices/sunspot-numbers/cycle-data/table\_cycle-dates\_maximum-minimum.txt)
 and ISN (www.sidc.be/silso/cyclesminmax) series, definitions of the cycle maxima and minima are not unambiguous
 and may involve information going beyond the sunspot counts \citep[see discussion in][]{hathawayLR}.
The often-used method of the 13-month running-mean smoothing sometimes leads to an unstable definition as it is affected by short
 bursts of activity \citep[e.g.,][]{carrasco_cycle_16}.
\begin{figure}[t]
\centering
\includegraphics[width=0.8\columnwidth]{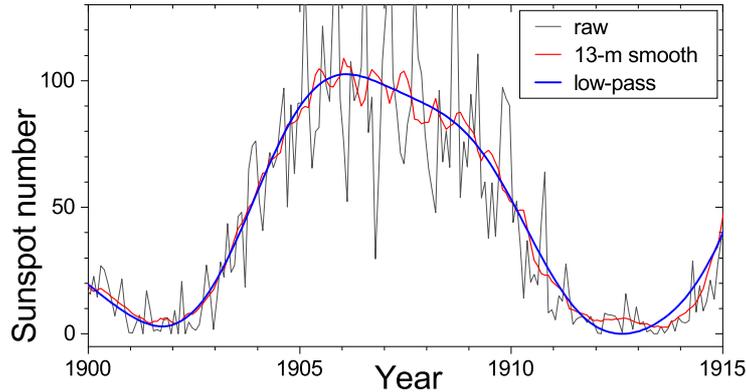}
\caption{Time series of the ISN-v2 for Solar Cycle 14.
 The original monthly sunspot numbers are shown with the thin black line, while 13-month unweighted running mean and the low-pass (0.25 yr$^{-1}$)
  filtered series are shown with the thick red and blue lines, respectively.}
\label{Fig:min}
\end{figure}

An example is shown in Figure~\ref{Fig:min} for Cycle 14 as presented in the ISN-v2 series.
One can see that the raw monthly data exhibit a strong variability up to 70\,\%, particularly around solar maximum.
A traditional way to define the dates of cycle maxima and minima refers to the extrema of the 13-month running
 mean, viz. the 13-month boxcar in the time domain with (the \textit{Gleissberg} filter) or without halving
 the weight for the edge point, as depicted by the red curve in Figure~\ref{Fig:min}.
However, this curve also depicts strong variability leading to partly ambiguous definitions, since the
 moving average is a very poor low-pass filter, due to its slow roll-off and poor stopband attenuation.
It has a broad frequency response with many sidelobes so that high-frequency variability leaks into
 the low-frequency range.
For example, the peak of the 13-month smoothed series in early 1906 corresponds to a deep drop in sunspot number.
The definition of the minima is not unambiguous either: the minimum of Cycle 14 can be either in mid-1901 or
 in early 1902 as set by random fluctuations of sunspot occurrence.
The minimum of Cycle 15 is even less clear as it can be early 1912 or late 1914.
\begin{table}[t]
\caption{Times (as fractional years) of the solar cycle minima and maxima for Cycles 1\,--\,22, defined as described
 in Section~\ref{S:cycle} for the four sunspot (group) series analyzed.
 Cycle numbering is the standard one.}
\label{Tab:min}
\begin{tabular}{r cc cc cc cc}
\hline
\#&\multicolumn{2}{c }{ISN-v1}&\multicolumn{2}{c }{ISN-v2}&\multicolumn{2}{c }{GSN-HS}&\multicolumn{2}{c}{GSN-ADF}\\
 & min & max & min & max & min & max & min & max\\
 \hline
1 & 1755.46 & 1761.38 & 1755.54 & 1761.38 & 1755.63 & 1760.46 & 1756.22 & 1760.63\\
2 & 1766.38 & 1770.05 & 1766.29 & 1770.04 & 1766.13 & 1770.05 & 1766.47 & 1769.88\\
3 & 1775.54 & 1778.88 & 1775.54 & 1778.88 & 1775.63 & 1779.38 & 1775.47 & 1778.97\\
4 & 1784.63 & 1788.13 & 1784.71 & 1788.13 & 1784.55 & 1788.13 & 1784.55 & 1788.05\\
5 & 1798.21 & 1803.71 & 1798.13 & 1803.71 & 1798.54 & 1802.05 & 1798.13 & 1802.38\\
6 & 1809.63 & 1816.46 & 1809.63 & 1816.46 & 1809.21 & 1817.29 & 1809.97 & 1815.88\\
7 & 1822.54 & 1830.29 & 1822.38 & 1830.38 & 1822.05 & 1830.29 & 1824.88 & 1830.22\\
8 & 1833.79 & 1837.29 & 1833.79 & 1837.29 & 1833.79 & 1837.21 & 1833.80 & 1837.22\\
9 & 1843.88 & 1848.46 & 1843.79 & 1848.63 & 1843.54 & 1848.88 & 1843.47 & 1848.97\\
10 & 1856.04 & 1860.38 & 1856.04 & 1860.29 & 1856.13 & 1860.46 & 1856.13 & 1860.47\\
11 & 1867.21 & 1871.04 & 1867.21 & 1871.04 & 1867.13 & 1871.13 & 1867.05 & 1871.05\\
12 & 1878.29 & 1883.54 & 1878.29 & 1883.54 & 1878.29 & 1883.96 & 1878.30 & 1883.97\\
13 & 1889.21 & 1893.79 & 1889.29 & 1893.71 & 1888.96 & 1893.88 & 1889.13 & 1893.88\\
14 & 1901.71 & 1906.12 & 1901.71 & 1906.13 & 1901.62 & 1906.20 & 1901.72 & 1906.22\\
15 & 1912.54 & 1917.87 & 1912.63 & 1917.88 & 1912.71 & 1917.87 & 1912.72 & 1917.80\\
16 & 1923.45 & 1927.62 & 1923.46 & 1927.71 & 1923.45 & 1927.54 & 1923.38 & 1927.63\\
17 & 1933.54 & 1938.04 & 1933.63 & 1938.04 & 1933.46 & 1938.04 & 1933.55 & 1937.97\\
18 & 1944.29 & 1948.12 & 1944.21 & 1948.04 & 1944.21 & 1948.04 & 1944.13 & 1948.13\\
19 & 1954.12 & 1958.12 & 1954.13 & 1958.13 & 1954.04 & 1958.20 & 1954.05 & 1958.22\\
20 & 1964.71 & 1969.04 & 1964.71 & 1969.04 & 1964.62 & 1969.12 & 1964.55 & 1969.05\\
21 & 1976.29 & 1980.37 & 1976.29 & 1980.46 & 1976.12 & 1980.29 & 1976.05 & 1980.22\\
22 & 1986.20 & 1990.29 & 1986.29 & 1990.38 & 1986.20 & 1990.62 & 1986.38 & 1990.47\\
\hline
\end{tabular}
\end{table}

Accordingly, we used a low-pass filter (rectangular boxcar in the frequency domain) with the cutoff frequency
 0.25 yr$^{-1}$, corresponding to a four-year period \citep[similar to][]{mursula01}.
This low-pass curve (blue in Figure~\ref{Fig:min}) is smooth, with cycle extrema defined unambiguously
 irrespectively to the high-frequency noise.
Henceforth, we use the low-pass filtered curves to define the cycle maxima and minima as
 listed in Table~\ref{Tab:min} for all the analyzed series.
The dates of the cycle minima are well constrained and agree within 4\,--\,5 months between series,
 with the sole exception of Cycle 7, where the difference between GSN-HS (1822.05) and GSN-ADF (1824.88)
 dates is 34 months.
The dates for the cycle maxima typically agree within six months between different series, except for Cycles
 5 (20 months between ISN and GSN-HS series) and 6 (17 months between GSN-ADF and GSN-HS series).

We accordingly re-defined the quantitative parameters of the solar cycle strength for each cycle and each analyzed series.
First, we considered the magnitude of the cycle [$S_{\rm max}$] as the maximum of the low-pass filtered curve,
 which gives a more robust value
 defined by the whole cycle evolution rather than by the very ``local'' activity bursts (see Figure~\ref{Fig:min}).
The values are collected in Table~\ref{Tab:max}.
\begin{table}[t]
\caption{Maximum smoothed sunspot (group) numbers $S_{\rm max}$ and the integral sum $I$
 (see Section~\ref{S:cycle} for the definition) for Cycles 1\,--\,22 for the four analyzed sunspot (group) series.
 Cycle numbering is the standard one.}
\label{Tab:max}
\begin{tabular}{r cc cc cc cc}
\hline
 &\multicolumn{2}{c }{ISN-v1}&\multicolumn{2}{c }{ISN-v2}&\multicolumn{2}{c }{GSN-HS}&\multicolumn{2}{c}{GSN-ADF}\\
\#& $S_{\rm max}$ & $I$ &$S_{\rm max}$ & $I$ &$S_{\rm max}$ & $I$ & $S_{\rm max}$ & $I$\\
\hline
1 & 73.1 & 5623 & 119.6 & 9361 & 4.6 & 359 & 6.6 & 533\\
2 & 108.0 & 6451 & 179.1 & 10762 & 8.3 & 509 & 9.4 & 638\\
3 & 149.6 & 7385 & 250.0 & 12325 & 6.2 & 383 & 11.6 & 706\\
4 & 142.6 & 10098 & 238.3 & 16807 & 7.8 & 596 & 10.7 & 846\\
5 & 50.2 & 3425 & 84.5 & 5714 & 3.2 & 192 & 3.5 & 282\\
6 & 44.4 & 2810 & 74.4 & 4672 & 2.2 & 154 & 5.7 & 446\\
7 & 66.4 & 4761 & 110.1 & 7936 & 4.9 & 342 & 5.0 & 358\\
8 & 135.6 & 7852 & 225.7 & 13077 & 8.9 & 504 & 8.0 & 479\\
9 & 118.3 & 8283 & 204.4 & 14786 & 7.0 & 534 & 7.1 & 569\\
10 & 98.9 & 6546 & 188.1 & 12437 & 7.0 & 456 & 7.8 & 523\\
11 & 127.9 & 7483 & 213.4 & 12504 & 7.9 & 478 & 9.6 & 621\\
12 & 67.7 & 4551 & 113.3 & 7597 & 4.9 & 328 & 5.7 & 409\\
13 & 88.5 & 5598 & 146.9 & 9322 & 7.8 & 483 & 8.2 & 554\\
14 & 62.0 & 4436 & 102.6 & 7400 & 5.2 & 381 & 5.3 & 409\\
15 & 93.5 & 5342 & 155.2 & 8900 & 8.4 & 491 & 7.8 & 474\\
16 & 77.1 & 4963 & 128.8 & 8277 & 6.9 & 464 & 6.9 & 465\\
17 & 117.6 & 7220 & 195.1 & 12011 & 10.2 & 641 & 10.3 & 653\\
18 & 152.4 & 9065 & 219.0 & 13268 & 12.0 & 717 & 11.7 & 720\\
19 & 206.4 & 11476 & 291.3 & 16274 & 15.6 & 869 & 14.3 & 826\\
20 & 111.9 & 8412 & 159.1 & 11940 & 8.7 & 690 & 8.5 & 679\\
21 & 163.0 & 9949 & 232.5 & 13932 & 13.0 & 829 & 12.0 & 784\\
22 & 163.9 & 9420 & 219.8 & 12673 & 13.3 & 769 & 12.2 & 730\\
\hline
\end{tabular}
\end{table}

We have also calculated the integral intensity (cumulative sum of monthly sunspot/group numbers from one cycle
 minimum to the next one) of
 solar cycles, using the different series, as summarized in Table~\ref{Tab:max}.

\section{Analysis of Empirical Relations}
\label{S:analysis}
Here we overview the empirical relations for the for analyzed series.

\subsection{Waldmeier Rule}
The \textit{classical} Waldmeier rule refers to a highly significant correlation between the length
 of the ascending phase (from minimum to maximum) [$T_{\rm A}$] and the magnitude of the cycle [$S_{\rm max}$].
Sometimes, a \textit{simplified} Waldmeier rule is used relating the magnitude of the cycle to its
 full length, but it is much weaker \citep{usoskin_SP03,hathawayLR}.
Also, a relation between the length of a cycle and the magnitude of the subsequent cycle (called the $n+1$ rule)
 was found to be strong and statistically significant \citep{solanki_krivova02} as expected from
 the dynamo theory \citep{charbonneau00,charbonneauLR}.

\begin{figure}[t]
\centering
\includegraphics[width=1\columnwidth]{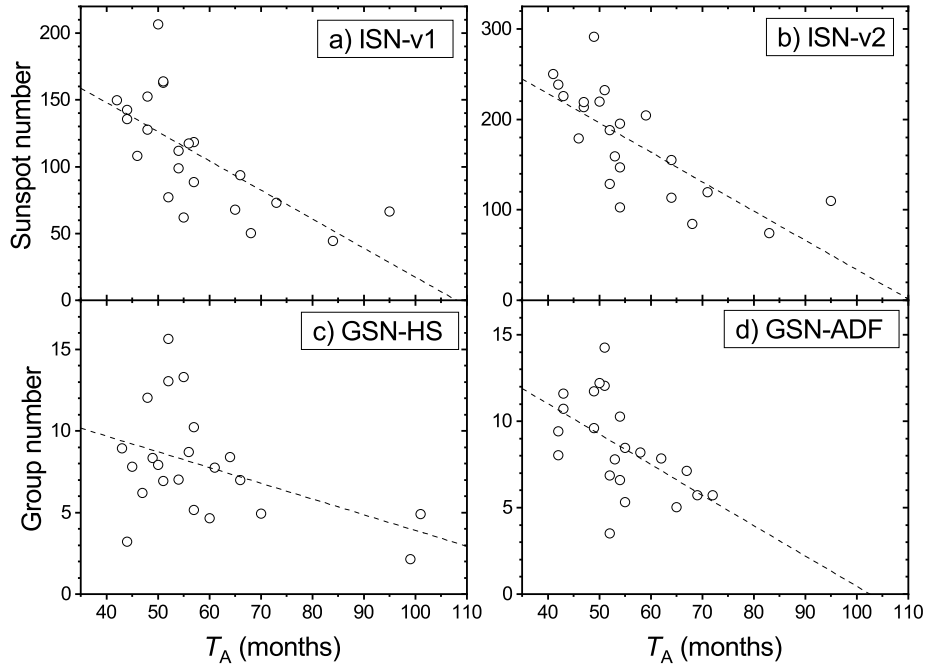}
\caption{Visualization of the classical Waldmeier rule with the scatter plots of the magnitude (Table~\ref{Tab:max}) vs.
 length of the ascending phase $T_{\rm A}$ (minimum to maximum) of solar cycles (individual dots) for the four
 analyzed sunspot series.
 The dotted lines depict the best-fit linear regression lines.
 Correlation coefficients and the significance are listed in Table~\ref{Tab:rule}  }
\label{Fig:Wald}
\end{figure}

Using the dates of the cycle minima and maxima (Table~\ref{Tab:min}) and magnitude [$S_{\rm max}$] (Table~\ref{Tab:max})
 we have estimated the Waldmeier rule for the four analysed series for Solar Cycles 1\,--\,22, using the three
 formulations described above.
The relations for the classical Waldmeier rule are depicted in Figure~\ref{Fig:Wald} and summarized in Table~\ref{Tab:rule}.

The length of the ascending phase varies substantially, ranging between 40 and 100 months.
It is interesting that weak Cycles 6 and 7 are characterised by exceptionally long ascending phases (80\,--\,100 months)
 in three of the analyzed series
 (Figure~\ref{Fig:Wald}a\,--\,c) probably affecting the overall Waldmeier rule.
On the other hand, the GSN-ADF exhibits a more compact spread of the $T_{\rm A}$ values (42\,--\,72 months),
 making Cycles 6 and 7 non-outliers.
\begin{table}[t]
\caption{The Pearson's linear correlation coefficients $r$ and their significance ($p-$values) corresponding to the
 Waldmeier rule in different formulations (see text) for the four analyzed series. }
\label{Tab:rule}
\begin{tabular}{cr cccc}
\hline
\multicolumn{2}{c|}{Waldmeier rule} & ISN-v1 & ISN-v2 & GSN-HS & GSN-ADF\\
\hline
Classical & $r$ &$-0.68_{-0.11}^{+0.14}$&$-0.73_{-0.09}^{+0.13}$&$-0.44_{-0.16}^{+0.20}$&$-0.54_{-0.14}^{+0.18}$\\
 & $p$ & $<0.001$ & $<0.001$ & 0.015 & 0.009\\
\hline
Simplified & $r$ &$-0.35_{-0.18}^{+0.22}$&$-0.29_{-0.20}^{+0.24}$&$-0.32_{-0.18}^{+0.22}$&$-0.27_{-0.20}^{+0.25}$\\
 & $p$ & 0.07 & 0.12 & 0.10 & 0.14\\
\hline
(n+1) & $r$ &$-0.66_{-0.12}^{+0.16}$&$-0.71_{-0.09}^{+0.13}$&$-0.42_{-0.17}^{+0.21}$&$-0.59_{-0.12}^{+0.18}$\\
 & $p$ & $<0.001$ & $<0.001$ & 0.015 & $<0.001$\\
\hline
\end{tabular}
\end{table}

It can be concluded that the Waldmeier rule appears stable against the exact data series.
The classical rule is found highly significant ($p<0.01$) for both versions of the ISN and GSN-ADF series,
 and significant ($p=0.015$) for the GSN-HS series.
The simplified rule is insignificant for all of the analyzed series ($p>0.05$).
The $(n+1)$ rule appears significant ($p=0.015$) for the GSN-HS series and highly significant ($p<0.01$)
 for other series.

\subsection{Gnevyshev--Ohl Rule}

\begin{figure}[t]
\centering
\includegraphics[width=1\columnwidth]{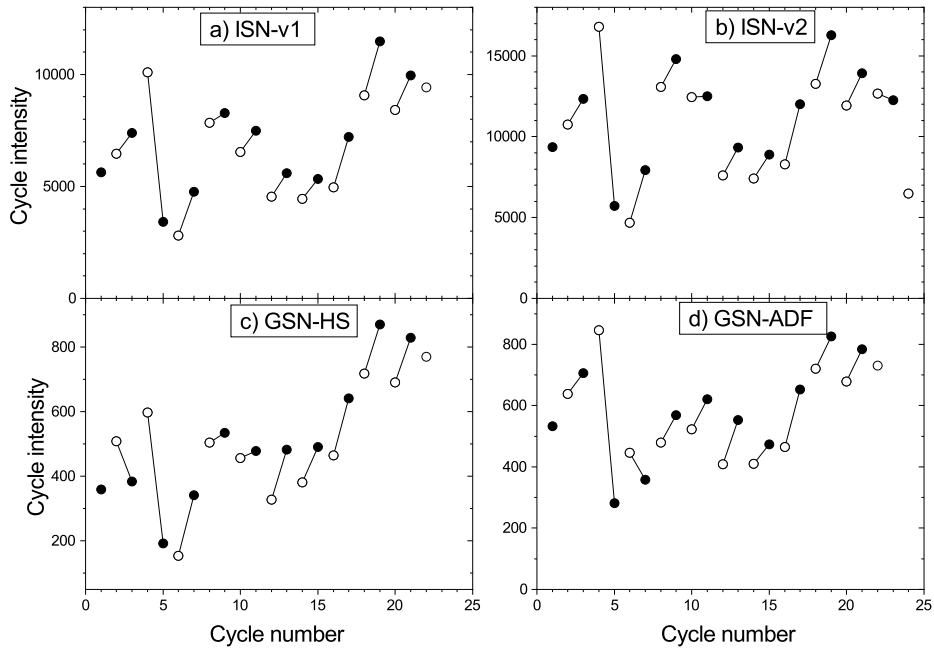}
\caption{Illustration of the GO rule for the four  series analyzed.
The $y$-axes values correspond to the cycle-integrated intensities (Table\ref{Tab:max}).
The open and filled circles denote the even and odd-numbered (in the standard numbering) cycles, respectively, connected in even-odd pairs.
The valid GO rule corresponds to the positively sloped line segments, while negatively sloped segments indicate breaks
 of the rule.}
\label{Fig:GO}
\end{figure}

The Gnevyshev-Ohl (GO) rule says that the solar cycles tend to be grouped in pairs of even- and the following
 odd-numbered cycles so that the odd cycle is stronger (in the sense of the integral cycle-accumulated intensity [$I$]) than
 the even one.
The numbering of the cycles follows the standard one introduced by R. Wolf (see Table~\ref{Tab:min}).
The GO rule for the four analyzed series is shown in Figure~\ref{Fig:GO}.

For the ``classical'' ISN-v1 series (Figure~\ref{Fig:GO}a), the GO rule is valid since the Dalton minimum (Cycles 6\,--\,21),
 but it breaks before that.
The revised ISN-v2 series (panel b) displays a similar pattern, viz. the GO rule breaks in the entrance to the Dalton minimum.
The GO rule also fails for the cycle pair 22\,--\,23.
Additionally, the GO rule is nearly broken for the cycle pair 10\,--\,11, when the odd cycle is stronger than the even cycle by only 0.5\,\%.
The GO rule of the GSN-HS series (panel c) is similar to that in the ISN-v1 series since the Dalton minimum but inverts the sign for the
 cycle pair 2\,--\,3.
The GSN-ADF series (panel d) is similar to other series, in the sense of the GO rule, but it exhibits an inverted sign for the
 cycle pair 6\,--\,7, during the Dalton minimum.
This may be related to the fact that the ADF method tends to overestimate solar activity during/around
 grand minima \citep{willamo18}.

In summary, the GO rule appears robust against the sunspot series used for the period after the Dalton minimum (after Cycle 8),
 while all series imply the breaking of the rule in the beginning of the Dalton minimum.
The ISN-v2 series leads to almost a break of the GO for Cycles 10\,--\,11.

The physical cause of the GO rule is not understood \citep{hathawayLR} and this empirical rule provides an observational constraint to the
  dynamo theory.
In particular, its interpretation in terms of the relic solar magnetic field \citep{mursula01} was disproved.
The break of the GO rule across the Dalton minimum can originate from three possible sources:
i) poor quality of data across the Dalton minimum with large uncertainties of the reconstructed sunspot (group) numbers,
 leading to the `lost' solar cycle in the very end of the 18th century \citep{usoskin_lost_AA_01,usoskin_arlt_09};
ii) phase catastrophe in solar cyclic evolution \citep[e.g.,][]{vitinsky86,kremliovsky94};
iii) random coincidence, since the GO is not a strict law but a statistical tendency.
However, the fact that the GO rule fails also for the cycle pair 22\,--\,23 \citep{komitov01} just prior to the
 modern low Cycle 24, favours the second option of the phase catastrophe accompanying a quick change of solar
 activity from a high to a low level.

\section{Conclusions}

We have tested the robustness of empirical rules related to the evolution of the 11-year Schwabe solar cycle
 against recent updates and modifications of the sunspot (group) series.
Four series with monthly temporal resolution have been analyzed for the period between 1749 and 1996 (Solar Cycles
  1\,--\,22): the classical Wolf sunspot number series (ISN-v1); the revised Intentional sunspot number series (ISN-v2);
 the original group sunspot number series (GSN-HS); and an updated sunspot group number based on the ADF method (GSN-ADF).
A more robust method (low-pass filter with the cutoff frequency of 0.25 year$^{-1}$) was used for definition of the cyclic evolution
 with respect to the traditional 13-month moving average method.

The \textit{Waldmeier rule} have been analyzed in three formulations (Table~\ref{Tab:rule}):
\begin{itemize}
\item\textit{Classical} formulation (the magnitude of a cycle is inversely related to the length of its ascending phase) appears
 significant for the GSN-HS series and highly significant for the other three series.
\item\textit{Simplified} formulation (the magnitude of a cycle is inversely related to its entire length) is insignificant
 for all the series.
\item\textit{(n+1)} formulation (the magnitude of a cycle is inversely related to the total length of the preceding cycle) was found
 significant for the GSN-HS series and highly significant for the other three series.
\end{itemize}
Thus, the Waldmeier rule in its classical and (n+1) formulations is robust and independent of the exact sunspot (group) series
 \citep[see also][]{aparicio12}.

The \textit{Gnevushev--Ohl rule} (solar cycles are paired in even--odd pairs so that the odd cycle is more intense than the preceding
 even cycle) was found robust for all series for Cycles 8\,--\,21 (although it nearly fails for Cycles 10\,--\,11 for the ISN-v2 series),
 but unstable across the Dalton minimum and before it.
The GO rule also breaks for the recent Cycles 22\,--\,23.

These empirical rules form important observational constraints for the solar dynamo theory \citep[e.g.,][]{charbonneauLR},
 but we leave aside theoretical interpretations focusing merely on the statistical results and their (in)dependence
 on the exact series used.
The result implies that the Waldmeier rules can be considered as very robust, while the Gnevyshev--Ohl rule
 is robust only after Solar Cycle 8 and becomes unstable before that.

\begin{acks}
This work was partially supported by the Academy of Finland (project No. 321882 ESPERA) and by the Russian Science Foundation
 (RSF project No. 20-67-46016).
The authors benefited from discussions within the ISSI International Team work \#417 (Recalibration of the Sunspot Number Series)
 and ISWAT-COSPAR S1-01 team.
\end{acks}


\appendix

\section{Update of the Sunspot Group Numbers Using the ADF method} 
    \label{App:A}
The active-day fraction (ADF) method of the sunspot (group) number reconstruction was invented by \citet{usoskin_ADF_16} and greatly
 upgraded by \citet{willamo17}.
It is based on the assumption that the `quality' of a solar observer is entirely defined by his/her observational
 threshold, rather than a scaling factor used in other methods.
An important advantage of the ADF method is that it calibrates the observers directly to the reference series
 avoiding the daisy-chain calibration intrinsic to other methods.
As a drawback, the ADF method tends to strongly overestimate the low-activity and moderately underestimate
 the high-activity cycles \citep{willamo18}.

After the publications of the ADF-based reconstruction by \citet{willamo17}, based on the sunspot group observation database
 of \citet{hoyt98}, an essential update was made with the upgrade and systematisation of all known sunspot records,
  as collected in a database of \citet{vaquero16}.
Accordingly, we have revisited the ADF-based sunspot group number reconstructed here, following exactly the algorithm
 described by \citet{willamo17}, but using the new sunspot number database \citep{vaquero16}.
As the reference observer, we used the Royal Greenwich Observatory (solarscience.msfc.nasa.gov/greenwch.shtml),
 similarly to \citet{willamo17}.

\begin{figure}[t]
\centering
\includegraphics[width=1\columnwidth]{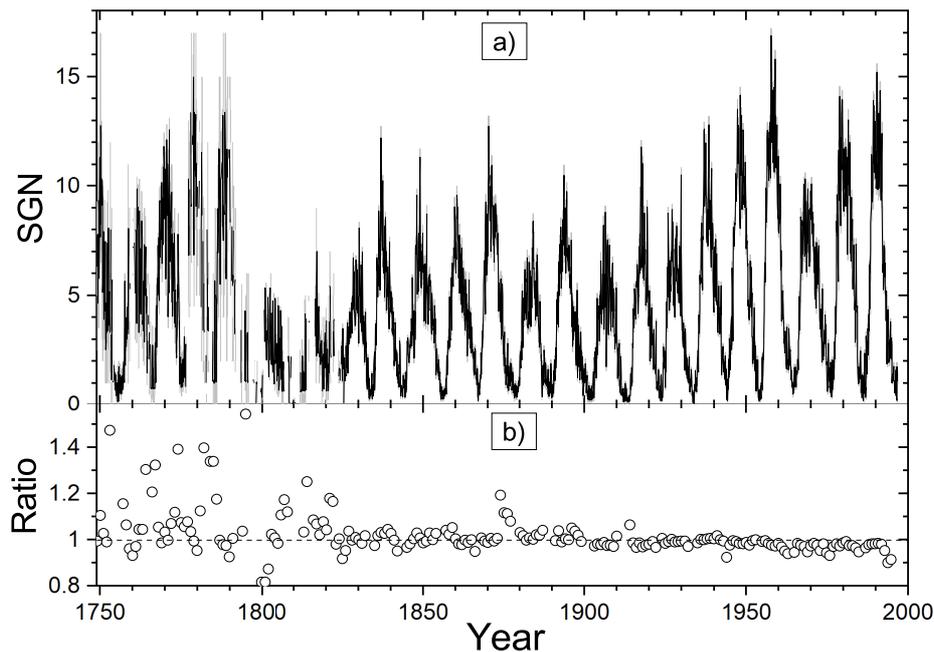}
\caption{a) Monthly sunspot group number reconstructed here using the ADF method and the updated solar observation
 database \citep{vaquero16}.
Uncertainties ($\pm 1\sigma$) are shown in grey.
Tabulated values are provided in the Electronic Supplementary Materials.
b) Ratio of the annual values of the SGN obtained here to those from \citet{willamo17}.
The unity value is shown by the dashed line.
The ratio values are not shown for the periods around solar cycle minima (SNG$<$1).
}
\label{Fig:ADF}
\end{figure}
The resultant data-series for the period of 1749\,--\,1996 is shown in Figure~\ref{Fig:ADF}a and tabulated in the
 Electronic Supplementary Materials as monthly and annual sunspot group series.
The ratio of the (annual) series obtained here to that of \citet{willamo17} is shown in Figure~\ref{Fig:ADF}b.
It is kept close (within 10\,\%) to unity after 1880, with small trembling being related to the Monte-Carlo procedure of the
 observers' calibration.
The ratio can, however, vary by up to 40\,\% before the Dalton minimum.
On the other hand, the ratio is consistent with the unity within the error bars.

This series supersedes the previous ADS-based SGN \citep{usoskin_ADF_16,willamo17} and is used here for further analysis.


\end{article}

\end{document}